\documentstyle[prc,aps,psfig]{revtex}

\begin{document}

\draft

\twocolumn[ \hsize\textwidth\columnwidth\hsize\csname
@twocolumnfalse\endcsname

\title{Faddeev approach to confined three-quark problems}

\author{Z.\ Papp$^{1}$, A.\ Krassnigg$^{2}$,  and W.\ Plessas$^{2}$ \\
$^1$ Institute of Nuclear Research of the 
Hungarian Academy of Sciences,
 H--4001 Debrecen, Hungary \\
$^2$ Institute for Theoretical Physics, University
of Graz,  A--8010 Graz, Austria }
\date{\today}
\maketitle

\begin{abstract}
We propose a method that allows for the efficient solution of the
three-body Faddeev equations in the presence of infinitely
rising confinement interactions. Such a method is 
useful in calculations of
nonrelativistic and especially semirelativistic constituent quark
models.
The convergence of the partial wave series is accelerated and possible
spurious contributions in the Faddeev components are avoided.
We demonstrate how the method works with the example of the 
Goldstone-boson-exchange chiral quark model for baryons.
\end{abstract}


\vspace{0.5cm} 

]

\narrowtext

\section{introduction}

The Faddeev approach has for a long time  been a very
efficient tool for solving three-body problems in nuclear physics.
Especially in the case of short-range forces  the Faddeev equations 
can nowadays
be solved to a high precision both for bound and scattering problems.
The Faddeev theory offers a number of advantages over other approaches,
such as solving the Schr\"odinger differential equation. In particular
one can
immediately include all relevant channels with the proper boundary
conditions; in the cases of either two or three identical particles, a
straightforward scheme for antisymmetrization is implemented; and
finally
a wide class of (local and nonlocal) two- and three-body interactions
can be
dealt with.

However, with the Faddeev integral equations one encounters
difficulties whenever the interactions are long-ranged.
There is the notorious problem of solving the
three-body Coulomb problem. Similarly, the solution of the constituent
quark
models poses practical difficulties. Irrespective of the type of
confinement,
one is confronted with complications due to the fact that the potentials
are
rising infinitely, as a function of inter-quark distance.

In the past the Faddeev theory has already been successfully applied to
the
calculation of nonrelativistic constituent quark models. In this case,
when
the confinement in general remains relatively weak, a direct solution of
the usual Faddeev equations can be achieved with a sufficient accuracy,
and a variety of results for baryon spectra has been obtained
\cite{a,b,c,d}.

The situation becomes much more difficult for semirelativistic
constituent 
quark models (i.e.\ with the kinetic-energy operator taken in
relativistic 
form); in this case the confinement is considerably enhanced, to a 
strength practically consistent with the string tension of quantum 
chromodynamics. The convergence of the partial wave series gets
slow, a relatively large number of channels needs to be
included, and one may also easily pick up spurious contributions in 
the Faddeev components, which get wiped out only in the total three-body
wave functions. The corresponding problems are, of course, already there
in a nonrelativistic framework and they were thoroughly studied
(e.g. in Ref.\ \cite{barnea}) for 
the case of the harmonic-oscillator potential, which sometimes 
is also used as a simplified model for confinement.

In the attempt to make the Faddeev approach directly amenable to any
three-body
problem with confining interactions, be it in a nonrelativistic or
semirelativistic framework, one of us has previously suggested
a different set-up of the Faddeev formalism \cite{pzconf}. The proposed 
method exploits the asymptotic filtering property of the Faddeev scheme
by
adopting a different splitting of the full three-body Hamiltonian.
Already in the nonrelativistic case considered in Ref.\ \cite{pzconf}
this
has led to a more efficient solution of the (modified) Faddeev
equations. 
Here we further elaborate on 
this method and demonstrate its efficiency even in the case
of semirelativistic constituent quark models.

Usually the full Hamiltonian $H$ of the system under consideration is
split into a free Hamiltonian (kinetic-energy operator) $H^0$ and an
interaction part $H^I$; the latter contains all two and three-body
forces.
However, this is not absolutely necessary. One may equally well adopt
a different splitting of $H$, where part of the interactions are put
into
a modified Hamiltonian $\widetilde{H}^0$ and the residual ones are
retained 
in $\widetilde{H}^I$. This turns out to be 
especially useful whenever there are
interactions that cause difficulties in the original Faddeev scheme. 
It is usually the case 
for interactions that do not fall off fast enough at large
distances, e.g., the Coulomb forces and notably also infinite 
confinement potentials. 

In the following section we outline the method that can 
treat confined three-body problems in an efficient and accurate manner. 
Beyond solving nonrelativistic problems we demonstrate how the method
can also be 
applied to semirelativistic constituent quark models, which rely on 
a kinetic-energy operator in relativistic form. In Section 3 we show 
how to solve the modified Faddeev equations, and we prove the 
good performance of the method in Section 4, taking the specific
examples of the Goldstone-boson-exchange (GBE) constituent quark
model in the nonrelativistic \cite{frascati}
and semirelativistic \cite{gppvw,gpvw} versions.

\section{Treatment of confinement in the Faddeev approach}

We start from the total Hamiltonian of a nonrelativistic or a 
semirelativistic three-quark system, which can be written as
\begin{equation}
H= H^0 + v_\alpha+ v_\beta + v_\gamma,
\label{H}
\end{equation}
where $H^0$ is the three-body kinetic-energy operator and 
$v_\delta=v_\delta^c + v_\delta^{hf}$ 
represents the  mutual quark-quark interactions containing both the
confinement 
($v_\delta^c$) and hyperfine ($v_\delta^{hf}$) potentials in the
subsystems
$\delta=\alpha,\beta,\gamma$. For the moment we leave out genuine 
three-quark forces, which, however, can easily be included into the
formalism and the method we follow. In the nonrelativistic case we may
express
the kinetic-energy operator by four equivalent
forms
\begin{eqnarray}
H^0 &= & \frac {p_\alpha^2}{2 \mu_\alpha} +\frac {q_\alpha^2}{2
M_\alpha}=
\frac {p_\beta^2}{2 \mu_\beta} +\frac {q_\beta^2}{2 M_\beta}=
\frac {p_\gamma^2}{2 \mu_\gamma} +\frac {q_\gamma^2}{2 M_\gamma}
\nonumber \\
&= & \sum_{i=1}^3 \frac {k_i^2}{2 m_i},
\label{H0nrel}
\end{eqnarray}
i.e.\ either through individual particle momenta $\vec{k}_i$ in the
center-of-mass system or in terms of
relative momenta $\vec{p}_\delta$ and $\vec{q}_\delta$
conjugate to the usual Jacobi coordinates $\vec{x}_\delta$ and
$\vec{y}_\delta$, respectively ($\delta=\alpha,\beta,\gamma$). In Eq.\
(\ref{H0nrel}), $m_i$ denotes the individual particle mass, $\mu_\delta$
the
reduced mass in the two-body subsystem $\delta$, and $M_\delta$ the
reduced mass of this subsystem with the third particle $\delta$.
In the semirelativistic case the kinetic-energy operator takes the form
\begin{equation}
H^0 = \sum_{i=1}^3 \sqrt{ k_i^2 + m_i^2},
\label{H0rel}
\end{equation}
where again $\vec{k}_i$ are the individual particle three-momenta in the
frame
with total three-momentum $\vec{P}=\sum_{i=1}^3  \vec{k}_i =0$.
We note that a Hamiltonian as in Eq.\ (\ref{H})
together with the relativistic kinetic-energy operator (\ref{H0rel})
represents an allowed mass operator in the point-form formalism of 
Poincar\'{e}-invariant quantum mechanics, irrespective of the dynamical
origin of the interactions \cite{klink}.

In the conventional Faddeev treatment the total Hamiltonian (\ref{H})
is first split into the free Hamiltonian $H^0$ and the interaction
Hamiltonian $H^I=v_\alpha + v_\beta + v_\gamma$. Then a decomposition of
the
total wave function 
\begin{equation}
|\Psi \rangle
=|\psi_{\alpha} \rangle+|\psi_{\beta} \rangle+|\psi_{\gamma} \rangle
\label{Psi}
\end{equation}
is carried out such that each of the three components satisfies the
Faddeev
integral equations of the type
\begin{equation}   
 |{\psi}_{\alpha} \rangle= G_\alpha (E) v_\alpha (
|{\psi}_{\beta}\rangle +  |{\psi}_{\gamma}\rangle )
\label{feq}
\end{equation}
with $\alpha,\beta,\gamma$ a cyclic permutation. Herein the so-called
channel resolvent
\begin{equation} 
 G_\alpha (E) = (E-H^0- v_\alpha)^{-1}
\label{G}
\end{equation}
occurs, which contains the interactions in subsystem $\alpha$ but is
otherwise 
characterized by a free motion of the third particle relative to the
two-body
subsystem. This scheme works well for short-range interactions and makes
it 
possible to easily incorporate the correct asymptotic behavior. 
Strictly speaking the standard Faddeev scheme applies only for
potentials
falling of fast enough at large distances. When the interactions are
long-ranged, however, there is never a free motion of the third particle
relative 
to the two-body subsystem. This makes it necessary to modify the Faddeev
formalism. Otherwise one risks unpleasant properties in the Faddeev
components. In particular, for infinitely rising potentials spurious
contributions are picked up and also the partial-wave series becomes
slowly
convergent.

One can circumvent these difficulties by modifying the decomposition
(\ref{Psi}) such that all the long-range potentials are included in a
modified
channel Green's operator. 
Specifically, in our case at least the long-range parts of the
confinement
interactions in all subsystems $\alpha,\beta,$ and $\gamma$ should be
included
in the modified channel resolvent \cite{pzconf}. 
One can attain this goal  
by adopting a different splitting of the total Hamiltonian into
\begin{equation}
H= H^c + \tilde{v}_\alpha+ \tilde{v}_\beta + \tilde{v}_\gamma,
\label{Ht}
\end{equation}
where
\begin{equation}
 H^c = H^0 + \tilde{v}_\alpha^{c} + \tilde{v}_\beta^{c} +
\tilde{v}_\gamma^{c} 
\label{Hc}
\end{equation}
contains, besides the kinetic energy, 
the long-range parts $\tilde{v}^c_\delta$ 
of the confining interactions  ${v}^c_\delta$ in all subsystems. 
The potentials $\tilde{v}_\delta$ are the
residual interactions containing the hyperfine potentials and the
short-range
parts of the confinement.

Based on Eqs.\  (\ref{Ht}) and (\ref{Hc}) we now decompose the total
wave
function into 
\begin{equation}
|\Psi \rangle
=|\tilde{\psi}_{\alpha} \rangle+|\tilde{\psi}_{\beta} \rangle+
|\tilde{\psi}_{\gamma} \rangle,
\label{Psit}
\end{equation}
where the modified Faddeev components are defined as
\begin{equation}
|\tilde{\psi}_{\alpha} \rangle = G^c(E) \tilde{v}_{\alpha}|\Psi \rangle
\label{psidef}
\end{equation}
with
\begin{equation}
 G^c(E) = (E-H^c)^{-1}.
\label{gc}
\end{equation}
They fulfill the integral equations
\begin{equation}   
 |\tilde{\psi}_{\alpha} \rangle= G^c_\alpha (E) \tilde{v}_\alpha (
|\tilde{\psi}_{\beta}\rangle +  |\tilde{\psi}_{\gamma}\rangle ),
\label{feqm}
\end{equation}
with $\alpha,\beta,\gamma$ again a cyclic permutation. 
As compared to Eq.\ (\ref{G}) the new channel resolvent gets modified to 
\begin{equation} 
 G^c_\alpha (E) = (E-H^c- \tilde{v}_\alpha)^{-1}\ .
\label{Gt}
\end{equation}
It exhibits just the desired property of including the long-range
confining
interactions in all subsystems $\alpha,\beta,\gamma$. Only the
short-range 
potential $\tilde{v}_\alpha$ remains in the modified Faddeev equations
(\ref{feqm}). Specifically, since now $G^c_\alpha$ contains also the
long-range
parts $\tilde{v}_\beta^c + \tilde{v}_\gamma^c$ of the 
confinement interactions in channels
$\beta$ and $\gamma$, the dependence of the component 
$|\tilde{\psi}_\alpha \rangle$
on the Jacobi coordinate $\vec{y}_\delta$ can never become a free
motion.
Rather the proper confinement-type asymptotic conditions are imposed on 
$|\tilde{\psi}_\alpha \rangle$. As a result, spurious contributions are
avoided
in the individual Faddeev components,
and at the same time the partial-wave expansion
converges much faster.

The splitting of the interactions in Eqs.\ (\ref{Ht}) and (\ref{Hc}) has
to be
done with care. In general, the interaction parts put into $H^c$ must
not
produce any bound state.
Otherwise the proper behavior of the Faddeev components 
$|\tilde{\psi}_\alpha \rangle$ would again be spoiled. 
Suppose the potentials contained in $H^c$ would produce bound states.
Then at the corresponding energies, the resolvent $G^c(E)$ would become
singular. Consequently, according to Eq.\ (\ref{psidef}), any large
Faddeev component $|\widetilde{\psi}_\alpha\rangle$ could be generated 
even if the full solution 
$|{\Psi} \rangle$ remains infinitesimally small.
Therefore, besides the true physical solutions
of the Hamiltonian $H$, Eqs.\ (\ref{feqm}) would also produce 
spurious solutions associated with the discrete 
eigenstates of the Hamiltonian $H^c$ \cite{yakovlev}. 
These spurious solutions would
occur for any $\tilde{v}_\alpha$, thus having no bearing for the
physical
spectrum of $H$. Of course, when adding up the three individual
Faddeev components these spurious solutions would cancel
out. However, they would cause numerical
instabilities in the practical calculations. Therefore they should be 
avoided by not allowing  $H^c$ to produce any bound states.

In the case of confinement interactions the above requirement 
cannot strictly be met, since even the longest-range parts of the
infinitely rising potential generate bound states. However, there is a
practical way out: one needs to eliminate the bound states generated by 
$H^c$ only in the region of physical interest. 
Outside that domain, i.e.\ reasonably far above the physical spectrum,
they do not matter. In practice, upon splitting the interactions in 
the Hamiltonian (\ref{Ht}) an auxiliary short-range 
potential is introduced with no effect on the physically interesting 
states. It only serves the
purpose of cutting off the confinement interaction at short and
intermediate 
distances thus avoiding low-lying bound states of $H^c$. In the solution
of the Faddeev equations below we take a Gaussian form for the auxiliary
potential as used already in Ref.\ \cite{pzconf}.

\section{Solution of the modified Faddeev Equations}

We solve Eqs.\ (\ref{feqm}) along the Coulomb-Sturmian (CS)
separable expansion approach \cite{pzwp}. 
For a two-body system in an angular momentum state $l$ one makes use
of the CS functions \cite{rotenberg} defined by
\begin{equation}
\langle r|n \rangle =\left[ \frac{n!}{(n+2l+1)!}\right]
^{1/2}(2br)^{l+1} \exp({-br}) L_n^{2l+1}(2br)  \label{basisr}
\end{equation}
or
\begin{equation}
\langle p|n \rangle 
 = \frac{(n+l+1) l! \sqrt{2 n!} }{
\sqrt{\pi (n+2l+1)!}}  
 \frac{b (4bp)^{l+1}}{(p^2+b^2)^{2l+2}}G_n^{l+1}
\left(\frac{p^2-b^2}{p^2+b^2}
\right)  \label{basisp}
\end{equation}
in configuration and momentum spaces,
respectively. Here, $L$ are the Laguerre and $G$ 
the Gegenbauer polynomials, and $b$ is a parameter.
The CS functions form a complete set 
\begin{equation}
{\bf {1}}=\lim\limits_{N\to \infty }\sum_{n=0}^N|
\widetilde{n}\rangle
\langle n|=\lim\limits_{N\to \infty }{\bf {1}}_N,  \label{unity}
\end{equation}
where, in configuration space,  $|\widetilde{n}\rangle$
is defined by 
$ \langle r|\widetilde{n}\rangle = \langle r|n \rangle/r$.

In three-body Hilbert space one extends the basis by defining the 
direct product 
\begin{equation}
| n \nu  \rangle_\alpha = \left\{ 
| n \rangle_\alpha \otimes | \nu
\rangle_\alpha \right\}, \ \ \ \ (n,\nu=0,1,2,\ldots),  
\label{cs3}
\end{equation}
where the states $| n  \rangle_\alpha$ and $| \nu  \rangle_\alpha$ are
associated with the Jacobi coordinates $x_\alpha$ and $y_\alpha$,
respectively.
The curly brackets stand for the angular-momentum coupling including
orbital angular momentum, spin and isospin.
The completeness relation now takes the form 
\begin{equation}
{\bf 1} =\lim\limits_{N\to\infty} \sum_{n,\nu=0}^N |
 \widetilde{n \nu } \rangle_\alpha \ \mbox{}_\alpha\langle 
{n \nu } | = \lim\limits_{N\to\infty} {\bf 1}_{N}^\alpha\ .
\end{equation}
It should be noted that we can introduce
three analogous bases, which belong to different fragmentations
$\alpha$, $\beta$, and $\gamma$.

In the practical solution of the modified Faddeev equations (\ref{feqm}) 
we terminate the sums in Eq.\ (\ref{cs3}) at some finite $N$.
Of course, the basis can always be made large enough to produce any
desired accuracy.
Thereby we provide for a separable expansion of the short-range
potential 
$\tilde{v}_\alpha$ in the three-body
Hilbert space
\begin{equation}
\tilde{v}_\alpha \approx {\bf 1}_N^\alpha \tilde{v}_\alpha 
{\bf 1}_N^\beta 
 =   \sum_{n,\nu ,n^{\prime },
\nu ^{\prime }=0}^N |\widetilde{ n\nu }\rangle_\alpha \;
\underline{\tilde{v}}_{\alpha \beta }
\;\mbox{}_\beta \langle \widetilde{n^{\prime }
\nu^{\prime }}|\ ,  \label{sepfe}
\end{equation}
where 
$\underline{\tilde{v}}_{\alpha \beta} =
\mbox{}_\alpha \langle n\nu  |
\tilde{v}_\alpha |n^{\prime }\nu^{\prime}\rangle_\beta $.
Inserting expression (\ref{sepfe}) into Eq.\ (\ref{feqm}) yields 
\begin{equation}
|\tilde{\psi}_\alpha \rangle = G_\alpha^c(E)[{\bf 1}_N^\alpha
\tilde{v}_\alpha 
{\bf 1}_N^\beta |\tilde{\psi}_\beta \rangle +
{\bf 1}_N^\alpha \tilde{v}_\alpha {\bf 1}_N^\gamma 
|\tilde{\psi}_\gamma \rangle ].
\label{feqsapp}
\end{equation}

By applying the CS states 
$\mbox{}_\alpha \langle \widetilde{n \nu}|$ 
from the left one obtains a linear system of homogeneous equations
for the coefficients of the Faddeev 
components $\underline{\tilde{\psi} }_{\alpha \; n \nu}=\mbox{}_\alpha
\langle 
\widetilde{n\nu  } |\tilde{\psi} _\alpha \rangle $: 
\begin{equation}
\{ [\underline{G}^c(E)]^{-1}-
\underline{\tilde{v}} \} \underline{\tilde{\psi}}=0.  \label{fep1}
\end{equation}
A unique solution exists if and only if
\begin{equation}
\det \{ [ \underline{G}^c (E) ]^{-1}-
\underline{\tilde{v}} \} = 0.
\label{det}
\end{equation}
This condition determines the energy eigenvalues.

The matrices $\underline{G}^c(E)$ and $\underline{\tilde{v}}$ 
have block structure and their matrix elements are given by
\begin{equation}
\underline{G}^c_{\alpha \; n \nu n' \nu'} = \delta_{\alpha \beta} \;
\mbox{}_\alpha \langle \widetilde{n\nu}|
G_\alpha^c(E)|\widetilde{
n^{\prime }\nu ^{\prime }}\rangle _\alpha\ 
\label{GG}
\end{equation}
and
\begin{equation}
\underline{\tilde{v}}_{\alpha \beta \; n \nu n' \nu'}  = 
(1-\delta_{\alpha \beta})
\mbox{}_\alpha \langle {n\nu}| \tilde{v}_{\alpha} | 
n^{\prime }\nu ^{\prime }\rangle _\beta\ .
\label{vm} 
\end{equation}

Notice that the matrix elements of the resolvent are needed only
between states of one and the same partition $\alpha$ whereas  
the matrix elements of the
potentials are always to be taken between states of different 
partitions $\alpha $ and $\beta $. While the latter 
can easily be evaluated (numerically) either in configuration or in
momentum space, the matrix elements $\underline{G}_\alpha^c$ are more
involved.
In Eq.\ (\ref{fep1}) the inverse of the matrix
$\underline{G}_\alpha^c(E)$
is needed and thus we make use of the approximation
\begin{equation}
 (\mbox{}_\alpha \langle \widetilde{n\nu}|
G_\alpha^c(E)|\widetilde{
n^{\prime }\nu ^{\prime }}\rangle_\alpha)^{-1} \approx 
\mbox{}_\alpha \langle {n\nu}|
(E-H^c-\tilde{v}_\alpha)|{ n^{\prime }\nu ^{\prime }} \rangle_\alpha\ ,
\label{inverse}
\end{equation}
i.e.\ the inverse of the matrix containing the matrix elements of an
operator 
is replaced by the matrix containing the 
matrix elements of the inverse operator.
Since $H^c$ contains all the confining interactions, 
this approximation does not modify the character
of the spectrum of $\underline{G}_\alpha^c$ and,
like the approximation in Eq.\ (\ref{sepfe}), it becomes
exact in the limit $N\to\infty$.
The matrix elements on the r.h.s.\ of Eq.\ (\ref{inverse}) can be
calculated in a straightforward manner, as the matrix element of $E$
and the ones of the nonrelativistic
kinetic-energy operator are known in closed analytical form. 
The potential matrix elements can easily be calculated numerically. 
In the case of the relativistic kinetic energy (\ref{H0rel})
the matrix elements of the corresponding operator are most conveniently
calculated numerically in momentum space.

\section{Performance of the method}

In order to demonstrate the efficiency of our method we present 
nonrelativistic and
semirelativistic three-quark calculations for light baryons.
For the quark-quark dynamics we choose the  
Goldstone-boson-exchange chiral quark model \cite{frascati,gppvw,gpvw}.
The total quark-quark interaction is thus given by 
\begin{equation}
v =v^c + v^\chi
\end{equation}
where the confinement is of the linear form
\begin{equation}
v^c (r) =V_0+C r\ .
\end{equation}
The chiral interaction is derived from Goldstone-boson exchange and is
represented by the spin-spin component of pseudoscalar meson exchange
(for the corresponding formulae see  Refs.\ \cite{frascati,gppvw,gpvw}). 
The parameters of the nonrelativistic GBE chiral
quark model can be found in Ref.\ \cite{frascati}, whereas the
semirelativistic version is specified in Refs.\ \cite{gppvw,gpvw}.
For the nonrelativistic version of this model the solutions have
previously
been obtained in good agreement by both the standard Faddeev approach
and the stochastic variational method (SVM) \cite{svm}, whereas for the
semirelativistic model only SVM solutions have existed up to now.
Here we are completing the solutions with the Faddeev
integral-equation results also for the semirelativistic version of the 
GBE quark model.

In the splitting of the short- and long-range potential parts in the 
confinement interaction we introduce an auxiliary potential in Gaussian
form
such that
\begin{equation}
\tilde{v}^c (r) =V_0+C r + a_0 \exp [-(r/r_0)^2]
\end{equation}
and 
\begin{equation}
\tilde{v}^\chi (\vec{r}) = v^\chi (\vec{r}) - a_0 \exp [-(r/r_0)^2]\ .
\end{equation}
As explained in the previous section, this guarantees that the bound
states of $H^c$ are shifted far beyond the 
energy range of physical interest (i.e. the light-baryon ground states
and
resonances). The parameters of the auxiliary potential have been taken
as 
$a_0=3\;\mbox{fm}^{-1}$ and $r_0=1\;\mbox{fm}$. By this choice of the
parameter
values any bound states of $H^c$ are avoided below $\approx 2$ GeV. The
values
of $a_0$ and $r_0$ also influence the rate of convergence but not the
final
(converged) results.

With the quark-quark interactions prepared in this way we can go ahead
and
calculate the $N$ and $\Delta$ spectra by solving Eq.\ (\ref{fep1}).
The results for the ground states and the excitation spectra are shown
in Tables I-VI for both the nonrelativistic and semirelativistic 
versions of the GBE quark model. We have also displayed the rate of
convergence with respect to the number of channels included. The 
convergence rate turns out to be essentially the same for the 
nonrelativistic and semirelativistic cases. It is practically not 
influenced by the strength of confinement, which is more than three 
times stronger in the latter case. One observes that only a few
angular-momentum states are needed to attain satisfactory convergence.
All this is essentially due to
the modified splitting of the Hamiltonian according to Eq.\ (\ref{Ht}).
The convergence is faster for the $\Delta$ ground state (Table V).
Also for its first positive- and negative-parity excitations 
only a rather small
number of channels need to be included. In case of the nucleon and its
resonances at most $10$ channels are necessary. For the separable
expansion of
the potentials, according to Eq.\ (\ref{sepfe}), about $20$ terms are
perfectly
sufficient. 

In all cases satisfactory agreement with the results obtained with
the SVM is reached. 
In the present work we have also
succeeded in treating semirelativistic constituent quark models, whose 
kinetic-energy operator is of the form (\ref{H0rel}). Technically the
calculation of the semirelativistic case is more involved,
as the matrix elements
of $H^0$ must also be calculated numerically. Otherwise, however, the
procedure
is the same as in the nonrelativistic case. 

Finally we remark that we can also calculate wave functions by putting
together
the solutions for the Faddeev components. The total wave
functions are then expressed in terms of CS functions. This fact
may facilitate their use in further applications.

\section{Summary}

We have proposed an efficient method for solving the three-body Faddeev
equations for the case of infinitely rising confinement interactions.
The method basically consists in splitting of the total Hamiltonian
in a way 
different from the usual approach: the long-range parts of the
interactions are
incorporated into a modified resolvent, which takes the place of the
usual free
resolvent. Consequently the Faddeev decomposition of the total
three-body
wave function is carried out in a
different manner, and one derives a set of modified Faddeev equations.
In solving 
them one can avoid the difficulties associated with long-range confining
interactions.

We have demonstrated the performance of our method in the case of the
Goldstone-boson-exchange chiral quark model both for its nonrelativistic
and
semirelativistic versions. We have obtained a rapid convergence with
the number of angular-momentum states included, irrespective of the 
strength of confinement. Our results agree perfectly with the ones
obtained before \cite{frascati,gppvw,gpvw}. In particular,  for the
semirelativistic case we have
now confirmed within the Faddeev approach the
results that have heretofore only been available from a variational
method. It is comfortable to see that the two solution methods, which 
are quite distinct in nature, produce exactly the same answers for the
light
baryon spectra.

\acknowledgements
This work was supported by the Austrian-Hungarian
Scientific-Technical Cooperation within
pro\-ject A-14/1998 and by Hungarian OTKA grants No. T026233 and
T029003.

\begin{table}
\[ \begin{array}{|c|cccc||c|c|c|c|c|c|}\hline
\multicolumn{5}{|c||}{\mbox{Channels}}&\multicolumn{2}{|c|}{\mbox{Nucleon}}&
\multicolumn{2}{|c|}{N^\ast(1440)}&\multicolumn{2}{|c|}{N^\ast(1710)} \\
\hline
\# &l_x&l_y&s&t&\mbox{sr}&\mbox{nr}&
\mbox{sr}&\mbox{nr}&\mbox{sr}&\mbox{nr}\\ \hline
1 & 0 & 0  & 0 & 0&1614&1514&1991&1785&2293&1978\\
2 & 0 & 0  & 1&  1&954&953&1471&1478&1911&1780\\
3 & 1 & 1  & 1 & 0&952&952&1469&1475&1907&1759\\
4 & 1 & 1  & 0 & 1&950&952&1466&1471&1793&1723\\
5 & 2 & 2  & 0 & 0&947&945&1464&1470&1792&1722\\
6 & 2 & 2  & 1 & 1&940&942&1460&1468&1782&1714\\
7 & 3 & 3  & 1 & 0&939&942&1459&1468&1779&1714\\
8 & 3 & 3  & 0 & 1&939&942&1459&1467&1776&1714\\
9 & 4 & 4  & 0 & 0&939&942&1459&1467&1776&1714\\
10& 4 & 4  & 1 & 1&939&942&1459&1467&1775&1714\\ \hline
\multicolumn{5}{|c||}{\mbox{SVM}}&939&939&1459&1465&1776&1712\\ \hline
\end{array}\]    
\caption{Masses of the nucleon ground state and the first two
positive-parity
excitations for the semirelativistic (sr) and the nonrelativistic (nr)
versions
of the pseudoscalar GBE chiral quark model. 
The convergence with respect to including an increasing number of
angular-momentum states $l_x,l_y$ is demonstrated; $s$ and $t$ denote
the
subsystem spin and isospin. In all cases, $N=20$ separable terms have
been
included into the expansion. A comparison to the results obtained with
the
stochastic variational method is given.}
\end{table}

\begin{table}
\[ \begin{array}{|c|cccc||c|c|}\hline
\multicolumn{5}{|c||}{\mbox{Channels}}&
\multicolumn{2}{|c|}{N^\ast(1535)-N^\ast(1520)}\\  \hline
\#  &l_x&l_y&s&t&
\rule{1.6em}{0cm}\mbox{sr}\rule{1.6em}{0cm}&\mbox{nr}\\ \hline
1 &  0&  1  & 0 & 0&2004&1685\\
2 &  0 & 1  & 1&  1&1859&1639\\
3 &  1 & 0  & 1 & 0&1716&1601\\
4 &  1 & 0  & 0 & 1&1538&1543\\
5 &  2 & 1  & 1 & 1&1529&1538\\
6 &  1 & 2  & 0 & 1&1525&1536\\
7 &  1 & 2  & 1 & 0&1521&1534\\
8 &  2 & 1  & 0 & 0&1521&1533\\
9 &  3 & 2  & 1 & 0&1519&1533\\
10 & 3 & 2  & 0 & 1&1518&1532\\
11 & 2 & 3  & 0 & 0&1518&1532\\
12 & 2 & 3  & 1 & 1&1516&1532\\ \hline
\multicolumn{5}{|c||}{\mbox{SVM}}&1519&1530\\ \hline
\end{array}\]    
\caption{Same as Table I but for the negative-parity excitations
$N^\ast(1535)-N^\ast(1520)$; these two states are degenerate in the 
pseudoscalar GBE quark model. }
\end{table}

\begin{table}
\[ \begin{array}{|c|cccc||c|c|}\hline
\multicolumn{5}{|c||}{\mbox{Channels}}&
\multicolumn{2}{|c|}{N^\ast(1650)-N^\ast(1700)-N^\ast(1675)}\\  \hline
\# &l_x&l_y&s&t&
\rule{3em}{0cm}\mbox{sr}\rule{3em}{0cm}&\mbox{nr}\\ \hline
1 & 0&  1  & 1 & 1&1963&1749\\
2 & 1 & 0  & 1&  0&1649&1656\\
3 & 1 & 2  & 1 & 0&1647&1655\\
4 & 2 & 1  & 1 & 1&1645&1655\\
5 & 2 & 3  & 1 & 1&1645&1655\\
6 & 3 & 2  & 1 & 0&1644&1655\\
7 & 3 & 4  & 1 & 0&1644&1655\\
8 & 4 & 3  & 1 & 1&1643&1655\\ \hline
\multicolumn{5}{|c||}{\mbox{SVM}}&1647&1652\\ \hline
\end{array}\]    
\caption{Same as Table I but for the negative-parity excitations
$N^\ast(1650)-N^\ast(1700)-N^\ast(1675)$; these three 
states are degenerate in the 
pseudoscalar GBE quark model.}
\end{table}

\begin{table}
\[ \begin{array}{|c|cccc||c|c|}\hline
\multicolumn{5}{|c||}{\mbox{Channels}}&
\multicolumn{2}{|c|}{N^\ast(1680)-N^\ast(1720)}\\  \hline
\# &l_x&l_y&s&t&
\rule{1.6em}{0cm}\mbox{sr}\rule{1.6em}{0cm}&\mbox{nr}\\ \hline
1 &  1&  1  & 1 & 0&2453&2039\\
2 &  1 & 1  & 0&  1&2316&1997\\
3 &  0 & 2  & 0 & 0&1966&1727\\
4 &  2 & 0  & 1 & 1&1787&1695\\
5 &  0 & 2  & 1 & 1&1776&1689\\
6 &  2 & 0  & 0 & 0&1743&1686\\
7 &  2 & 2  & 0 & 0&1742&1686\\
8 &  2 & 2  & 1 & 1&1736&1685\\
9 &  3 & 1  & 0 & 1&1734&1684\\
10 & 3 & 1  & 1 & 0&1731&1683\\
11 & 1 & 3  & 0 & 1&1730&1683\\
12 & 1 & 3  & 1 & 0&1729&1682\\ \hline
\multicolumn{5}{|c||}{\mbox{SVM}}&1729&1680\\ \hline
\end{array}\]    
\caption{Same as Table I but for the negative-parity excitations
$N^\ast(1680)-N^\ast(1720)$; these two
states are degenerate in the 
pseudoscalar GBE quark model.}
\end{table}

\begin{table}
\[ \begin{array}{|c|cccc||c|c|c|c|}\hline
\multicolumn{5}{|c||}{\mbox{Channels}}&\multicolumn{2}{|c|}{\Delta(1232)}&
\multicolumn{2}{|c|}{\Delta^\ast(1600)}\\  \hline
\# &l_x&l_y&s&t&\mbox{sr}&\mbox{nr}&\mbox{sr}&\mbox{nr}\\ \hline
1 &0 &0  & 1 & 1&1244&1249&1721&1590\\
2 &2 &2  & 1 & 1&1239&1238&1718&1588\\
3 &4 &4  & 1 & 1&1239&1237&1718&1588\\
4 &6 &6  & 1 & 1&1239&1237&1718&1588\\
5 &8 &8  & 1 & 1&1239&1237&1718&1588\\ \hline
\multicolumn{5}{|c||}{\mbox{SVM}}&1240&1233&1718&1586\\ \hline
\end{array}\]
\caption{Same as Table I but for the $\Delta$ ground state  and its
first
positive-parity excitation.}
\end{table}

\begin{table}
\[ \begin{array}{|c|cccc||c|c|}\hline
\multicolumn{5}{|c||}{\mbox{Channels}}&
\multicolumn{2}{|c|}{\Delta^\ast(1620)-\Delta^\ast(1700)}\\  \hline
\# &l_x&l_y&s&t&\rule{1.6em}{0cm}
\mbox{sr}\rule{1.6em}{0cm}&\mbox{nr}\\ \hline
1 & 0&  1  & 1 & 1&1963&1749\\
2 & 1 & 0  & 0&  1&1644&1646\\
3 & 1 & 2  & 0 & 1&1642&1645\\
4 & 2 & 1  & 1 & 1&1640&1645\\
5 & 3 & 2  & 0 & 1&1639&1644\\
6 & 2 & 3  & 1 & 1&1638&1644\\
7 & 3 & 4  & 0 & 1&1638&1644\\
8 & 4 & 3  & 1 & 1&1638&1644\\ \hline
\multicolumn{5}{|c||}{\mbox{SVM}}&1642&1642\\ \hline
\end{array}\]    
\caption{Same as Table V but for the negative-parity excitations
$\Delta^\ast(1620)-\Delta^\ast(1700)$; these two
states are degenerate in the 
pseudoscalar GBE quark model.}
\end{table}


\begin{references}

\bibitem{a} B.\ Silvestre-Brac and C.\ Gignoux, Phys.\ Rev.\ D {\bf 32},
743
(1985).

\bibitem{b} C.\ Roux and B.\ Silvestre-Brac,
 Few-Body Systems {\bf 19}, 1 (1995).

\bibitem{c} B.\ Silvestre-Brac,
 Few-Body Systems {\bf 20}, 1 (1996).
 
\bibitem{d} L.\ Ya.\ Glozman, Z.\ Papp, W.\ Plessas, 
 Phys.\ Lett.\ {\bf B381}, 311 (1996).

\bibitem{barnea} N.\ Barnea and V.\ Mandelzweig, Phys.\ Rev.\ C {\bf
45},
1458 (1992); ibid. {\bf 49}, 2910 (1994).

\bibitem{pzconf} Z.\ Papp, Few-Body Systems {\bf 26}, 99 (1999)

\bibitem{frascati} L.\ Ya.\ Glozman, Z.\ Papp, W.\ Plessas, 
K.\ Varga, and R.\ F.\ Wagenbrunn,  
Nucl.\ Phys.\ {\bf A263}, 90c  (1997).

\bibitem{gppvw} L.\ Ya.\ Glozman, Z.\ Papp, W.\ Plessas, 
K.\ Varga, and R.\ F.\ Wagenbrunn, Phys.\ Rev.\ C {\bf 57}, 3406 (1998). 

\bibitem{gpvw} L.\ Ya.\ Glozman, W.\ Plessas, K.\ Varga, 
 and R.\ F.\ Wagenbrunn,  Phys.\ Rev.\ D {\bf 58}, 094030 (1998).

\bibitem{klink} W.\ H.\ Klink and M.\ Rogers, Phys.\ Rev.\ C {\bf 58},
3605
(1998); W.\ H.\ Klink, Phys.\ Rev.\ C {\bf 58}, 3617 (1998).

\bibitem{yakovlev} S.\ L.\ Yakovlev, Theor. Math. Phys. {\bf 102}, 235
(1995);
ibid. {\bf 107}, 835 (1996);  Few-Body Systems Suppl. {\bf 10}, 85
(1999).

\bibitem{pzwp}  Z.\ Papp and  W.\ Plessas, Phys.\ Rev.\ C 
{\bf 54}, 50 (1996).

\bibitem{rotenberg} M.\ Rotenberg, Ann.\ Phys.\  (N.Y.) {\bf 19}, 
262 (1962);  Adv.\ Atom.\ Mol.\ Phys.\  {\bf 6}, 233 (1970).

\bibitem{svm} Y.\ Suzuki and K.\ Varga, {\it Stochastic Variational
Approach to
Quantum-Mechanical Few-Body Problems} (Springer, Berlin, 1998).

\end{references}
\end{document}